\documentclass[
 aip,
 amsmath,amssymb,
 reprint,%
]{revtex4-1}

\usepackage{graphicx}
\usepackage{dcolumn}
\usepackage{bm}
\usepackage{xcolor}
\usepackage{soul}  
\usepackage{lineno}
\usepackage{siunitx} 
\usepackage[utf8]{inputenc}
\usepackage[T1]{fontenc}
\usepackage{mathptmx}
\usepackage[version=4]{mhchem}
\usepackage{comment}
\makeatletter
\def\@email#1#2{%
 \endgroup
 \patchcmd{\titleblock@produce}
  {\frontmatter@RRAPformat}
  {\frontmatter@RRAPformat{\produce@RRAP{*#1\href{mailto:#2}{#2}}}\frontmatter@RRAPformat}
  {}{}
}%
\makeatother
\begin{document}

\preprint{AIP/123-QED}

\title[Mid-Infrared Ring Interband Cascade Laser]{Mid-Infrared Ring Interband Cascade Laser: Operation at the Standard Quantum Limit}

\author{Georg Marschick}
\altaffiliation[]{These authors equally contributed to this work.}
 \affiliation{TU Wien -- Institute of Solid State Electronics \& Center for Micro- and Nanostructures, Gußhausstraße 25-25a -- 1040 Vienna, Austria.}
 
\author{Jacopo Pelini}%
\altaffiliation[]{These authors equally contributed to this work.}
\affiliation{University of Naples Federico II, Corso Umberto I 40 -- 80138 Napoli, Italy}
\affiliation{CNR-INO -- Istituto Nazionale di Ottica, Largo Fermi, 6 -- 50125 Firenze FI, Italy}
\author{Tecla Gabbrielli}
\author{Francesco Cappelli}
\affiliation{CNR-INO -- Istituto Nazionale di Ottica, Via Carrara, 1 -- 50019 Sesto Fiorentino FI, Italy}
\affiliation{LENS -- European Laboratory for Non-Linear Spectroscopy, Via Carrara, 1 -- 50019 Sesto Fiorentino FI, Italy}
\author{Robert Weih}
\affiliation{Nanoplus Nanosystems and Technologies GmbH, Oberer Kirschberg 4 -- 97218 Gerbrunn, Germany}
\author{Hedwig Knötig}
\affiliation{TU Wien -- Institute of Solid State Electronics \& Center for Micro- and Nanostructures, Gußhausstraße 25-25a -- 1040 Vienna, Austria.}
\author{Johannes Koeth}
\affiliation{Nanoplus Nanosystems and Technologies GmbH, Oberer Kirschberg 4 -- 97218 Gerbrunn, Germany}
\author{Sven Höfling}
\affiliation{Julius-Maximilians-Universität Würzburg -- Physikalisches Institut, Lehrstuhl für Technische Physik, Am Hubland -- 97074 Würzburg, Germany}
\author{Paolo De Natale}
\affiliation{CNR-INO -- Istituto Nazionale di Ottica, Largo Fermi, 6 -- 50125 Firenze FI, Italy}
\affiliation{CNR-INO -- Istituto Nazionale di Ottica, Via Carrara, 1 -- 50019 Sesto Fiorentino FI, Italy}
\affiliation{LENS -- European Laboratory for Non-Linear Spectroscopy, Via Carrara, 1 -- 50019 Sesto Fiorentino FI, Italy}
\affiliation{INFN -- Istituto Nazionale di Fisica Nucleare, Via Sansone, 1 -- 50019 Sesto Fiorentino FI, Italy}
\author{Gottfried Strasser}
\affiliation{TU Wien -- Institute of Solid State Electronics \& Center for Micro- and Nanostructures, Gußhausstraße 25-25a -- 1040 Vienna, Austria.}
\author{Simone Borri}
\affiliation{CNR-INO -- Istituto Nazionale di Ottica, Via Carrara, 1 -- 50019 Sesto Fiorentino FI, Italy}
\affiliation{LENS -- European Laboratory for Non-Linear Spectroscopy, Via Carrara, 1 -- 50019 Sesto Fiorentino FI, Italy}
\affiliation{INFN -- Istituto Nazionale di Fisica Nucleare, Via Sansone, 1 -- 50019 Sesto Fiorentino FI, Italy}
\author{Borislav Hinkov}
\affiliation{TU Wien -- Institute of Solid State Electronics \& Center for Micro- and Nanostructures, Gußhausstraße 25-25a -- 1040 Vienna, Austria.}
\email{borislav.hinkov@tuwien.ac.at}
\email{francesco.cappelli@ino.cnr.it}

\date{\today}

\begin{abstract}
Many precision applications in the mid-infrared spectral range have strong constraints based on quantum effects that are expressed in particular noise characteristics. They limit, e.g., sensitivity and resolution of mid-infrared imaging and spectroscopic systems as well as the bit-error rate in optical free-space communication. Interband cascade lasers (ICLs) are a class of mid-infrared laser exploiting interband transitions in type-II band alignment geometry. They are currently gaining significant importance for mid-infrared applications from $<$3~$\mu$m to $>$6~$\mu$m wavelength, enabled by novel types of high-performance ICLs such as ring-cavity devices. Their noise-behavior is an important feature that still needs to be thoroughly analyzed, including its potential reduction with respect to the shot noise limit. \\
In this work, we provide a comprehensive characterization of $\mathrm{\lambda} = \SI{3.8}{\um}$-emitting, continuous-wave ring-ICLs operating at room temperature. It is based on an in-depth study of their main physical intensity noise features, such as their bias-dependent intensity noise power spectral density (INPSD) and relative intensity noise (RIN). We obtain shot-noise-limited statistics for Fourier frequencies above 100~kHz. This is an important result for precision applications, e.g. interferometry or advanced spectroscopy, which benefit from exploiting the advantage of using such a shot-noise limited source, enhancing the setup sensitivity. Moreover, it is an important feature for novel quantum optics schemes including testing specific light states below the shot noise level, such as squeezed states.

\end{abstract}

\maketitle

\section{Introduction}

Interband cascade lasers (ICLs) are semiconductor-based, coherent mid-IR light sources, first demonstrated by Yang et al. in 1995~\cite{Yang1995}. They are the interband counterpart to quantum cascade lasers (QCLs), which instead rely on intersubband transitions~\cite{Faist2013a} and have been the dominant mid-IR lasers since their realization in 1994~\cite{Faist1994}. These sources have immediately attracted a wide interest in view of the many potential applications, with a focus on molecular species detection in solid~\cite{Fuchs2010,Amrania2018}, liquid~\cite{Mizaikoff2013,Dabrowska2022} and gas-phase~\cite{Griffiths2007,Kosterev2008}. This has sparked, e.g., important works in greenhouse gas detection of methane, carbon dioxide or nitrous oxide, including the detection of the most elusive gas-isotopes~\cite{Tuzson2013}, in high-sensitivity gas measurements down to the ppq-level~\cite{Galli2016ppq,DelliSanti2021} even in real-world applications, or in broadband ($>$10~cm$^{-1}$), high-resolution (MHz-range) spectroscopy techniques like dual-comb spectroscopy~\cite{Villares2014}. Moreover, other important mid-IR applications are currently getting significant attention, such as spectral imaging~\cite{Fuchs2010,Amrania2018,Razeghi2020} and, in more recent years, optical free-space communication~\cite{Dely2022,Pang2022,Flannigan2022}. \\
This large interest acted as a strong driving force for the technological development of these sources. ICLs differ from QCLs, e.g. by their much lower power consumption and their operation at shorter wavelengths, even below 3~$\mu$m. Due to these and other peculiarities, ICLs are nowadays in many fields competitive with their QCL counterparts, matching the requests for high optical output power~\cite{Vurgaftman2011a,Knotig2022}, wide spectral tunability~\cite{Shim2021}, comb emission~\cite{Sterczewski2020}, compact dimensions and integrability~\cite{Shim2021}, spectral control and (ultra-)narrow linewidth~\cite{Weih2015,knötig-ringICL,Borri2020} and low noise emission~\cite{Deng:19}. \\
ICLs are the result of combining the strong interband transitions and long recombination lifetimes inherent to traditional diode lasers~\cite{vurgaftman:2015} with the voltage-efficient cascading design of QCLs~\cite{Faist1994} into an active region (AR) using type-II band alignment. This allows maintaining the QCL-like flexibility in designing the emission wavelength of ICLs through band-structure engineering, while, simultaneously, strongly reducing their number of AR periods. As mentioned, the result is a significantly lower power consumption, e.g. at laser threshold~\cite{vurgaftman2013interband} of around 170~mW~\cite{knötig-ringICL}, to compare to even specifically optimized low dissipation QCLs with threshold dissipation values between 350--850~mW~\cite{Hinkov2012,FengXie2012,Bismuto2015}. This advantage is particularly important for portable ICL-based sensors~\cite{weih2013} or for future space deployment. In novel ring-ICLs, ring-shaped ridge cavities are used together with vertical light emission, merging multiple advantages into a single device. First, the ring-cavity shares a similar geometry as most discrete optical elements such as lenses and mirrors, being beneficial for light collimation or focusing. Second, the large effective surface area of circular waveguides offers a large aperture, providing small divergent emission beams with angles below $\pm 10^{\circ}$ and thus simpler collimation~\cite{knötig-ringICL}. Third, a previous work in QCLs has revealed that ring geometries, due to their different mode-distribution within the cavity as compared to straight ridges, offer specific, electronically controllable frequency-modulation (FM) states~\cite{Hinkov2019a}, which are useful features for high-speed spectroscopy~\cite{Hangauer2013}. Fourth, compared to other vertical surface-emitting lasers, such as vertical-cavity surface-emitting lasers (VCSELs)~\cite{vcsel,vcsel1}, with their limited output power due to small gain volumes, the output power of ring-ICLs can be scaled up by simply increasing ring diameter or the waveguide width. In this case, obeying to certain design guidelines prevents higher-order lateral modes~\cite{knötig-ringICL}. \\
For controlling linewidth, single-mode emission capabilities and vertical light outcoupling in ring-ICLs, distributed feedback (DFB) gratings in the laser cavity can be used which periodically modulate the complex refractive index of the waveguide following the Bragg condition.~\cite{Joannopoulos2008,Suess2016} DFB gratings have already been successfully integrated into ICLs using various waveguide geometries.\cite{Weih2015,holzbauer-ringICL,knötig-ringICL} For efficient vertical light extraction in ring-cavities, gratings following the 2\textsuperscript{nd}-order Bragg condition are used~\cite{holzbauer-ringICL,knötig-ringICL}, which combine vertical with single-mode emission. Moreover, this opens the pathway to implement 2D multi-wavelength concentric array geometries~\cite{Marschick2023}, an important step towards broadband chip-scale spectrometers. \\
Despite all the achievements of ICLs, their intensity noise together with its potential reduction in ring-ICLs still needs to be thoroughly characterized. Fundamentally, intensity noise in semiconductor lasers like ICLs originates from their various internal electronical and optical processes such as spontaneous emission and random carrier generation/recombination.~\cite{Deng:19} Understanding its characteristics is important for increasing the sensitivity and resolution of imaging or spectroscopic systems~\cite{Deng:19,Hodgkinson2013} and for telecommunication concepts with reduced bit-error rate~\cite{Tyson2002}. Furthermore, it is of particular relevance in the future development of quantum optics schemes, such as homodyne detection, where a shot-noise limited source is highly desirable, as local oscillator, to test light states below the shot noise level (e.g. squeezed states)~\cite{loudon:2000book,Gabbrielli:2021}. \\
In the current work, we follow this need and investigate for the first time the relative intensity noise (RIN) of a single-mode-emitting ring-ICL. The device  operates in continuous-wave (CW) mode at room temperature with an emission wavelength $\mathrm{\lambda}$~=~\SI{3.8}{\um}. As previously discussed, ring devices have beneficial features for spectroscopic applications as compared to similar ridge devices~\cite{knötig-ringICL,Hinkov2019a,Marschick2023}. In our study, we first analyze the light-current-voltage (LIV) and single-mode emission characteristics of a typical custom-made 2\textsuperscript{nd}-order DFB ring-ICL. Then, a balanced-detection setup, consisting of a 50/50 beam splitter and two identical photovoltaic detectors, is employed to characterize the intensity noise power spectral density (INPSD) of the ring-ICL and compare it to the directly measured shot noise level. We further analyze the RIN of the ICL under different laser driving conditions to understand the optimal low-intensity-noise working regime of the tested device geometry.

\section{Device structure and working principle}

\begin{figure}[!h]
    \centering
    \includegraphics[width=0.9 \columnwidth]{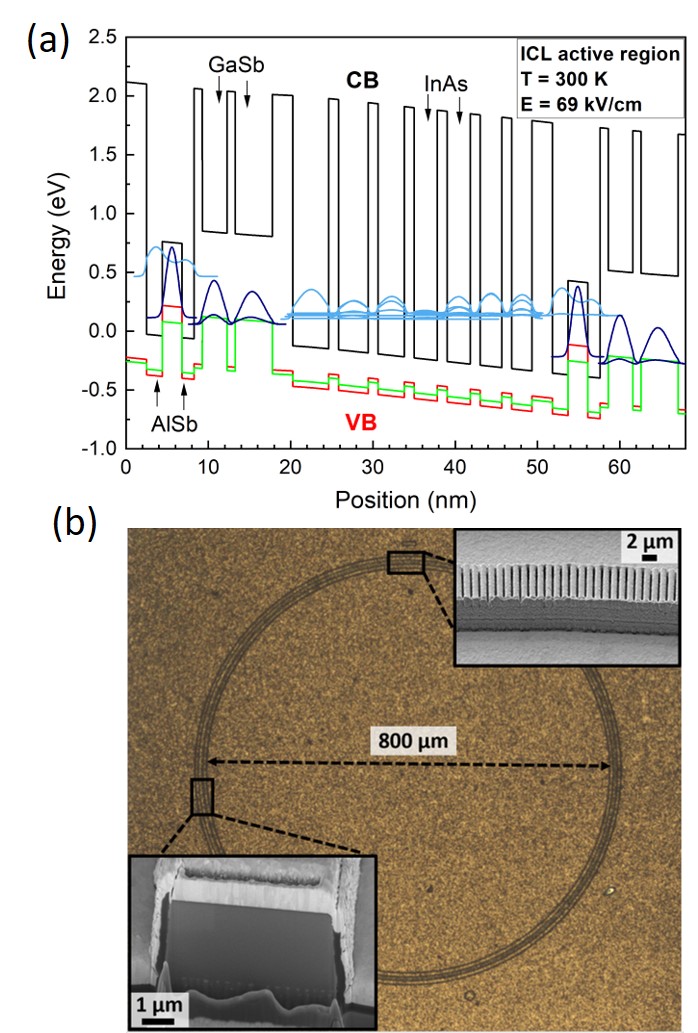}
    \caption{(a) Bandstructure of the ring-ICL including the simulated wavefunctions for an applied field of 69 kV/cm. (b) Microscope image of the fabricated ring-ICL. The insets show: (top right) a detailed view of the implemented 2\textsuperscript{nd}-order DFB grating for vertical light coupling and single mode emission and (bottom left) a scanning electron microscope (SEM) image of a focused ion beam (FIB) cut through the ridge of the ring device.}
    \label{fig:ICL_design}
\end{figure}

\noindent The quantum structure of the device investigated in this work is grown by solid-source molecular beam epitaxy (MBE) on n-GaSb (100) substrate. The w-shaped AR with 6 periods follows the layer sequence \SI{2.50}{\nm} \ce{AlSb} / \SI{1.92}{\nm} \ce{InAs} / \SI{2.40}{\nm} \ce{In_{0.35}Ga_{0.65}Sb} / \SI{1.49}{\nm} \ce{InAs} / \SI{1.0}{\nm} \ce{AlSb} for a target emission wavelength of \SI{3.8}{\um}. It is sandwiched between two 200~nm thick \ce{GaSb} separate confinement layers as well as a \SI{3.5}{\um} and \SI{2.0}{\um} \ce{InAs/AlSb} lower and upper cladding, respectively. Figure~\ref{fig:ICL_design}(a) shows the bandstructure including simulated wavefunctions of the AR design for an applied external field of 69 kV/cm. 
The epitaxial ICL structure is processed into ring-shaped cavities with a diameter of $\sim$800~µm and a ridge width of 5~µm (circumference: approximately 2.5~mm) using state-of-the-art cleanroom fabrication techniques. Special attention is given to the below 1-micrometer feature size of the implemented 2\textsuperscript{nd}-order DFB grating for vertical and single-mode light-coupling, which was realized using electron-beam lithography combined with reactive ion etching. Figure \ref{fig:ICL_design}(b) displays the microscope image of a typical finalized ring-ICL device, including scanning electron microscope (SEM) images of the DFB grating and a focused ion beam (FIB) cut through the ridge of the ring device revealing its high-quality cross-section profile. \\
Substrate-side emission is the preferred geometry for such devices, allowing to cover the entire topside of the rings including the DFB grating structure with gold and to use flip-chip bonding on copper submounts with indium solder. This results in significantly improved heat extraction from the device AR and is important for high-performance CW operation. More details on the AR design and device fabrication can be found elsewhere.\cite{knötig-ringICL}

\section{Device characterization}

\subsection{LIV curve and emission spectra characterization}
\label{subsex:livcurve-OS}
\begin{figure}[!h]
    \centering
    \includegraphics[width = \columnwidth]{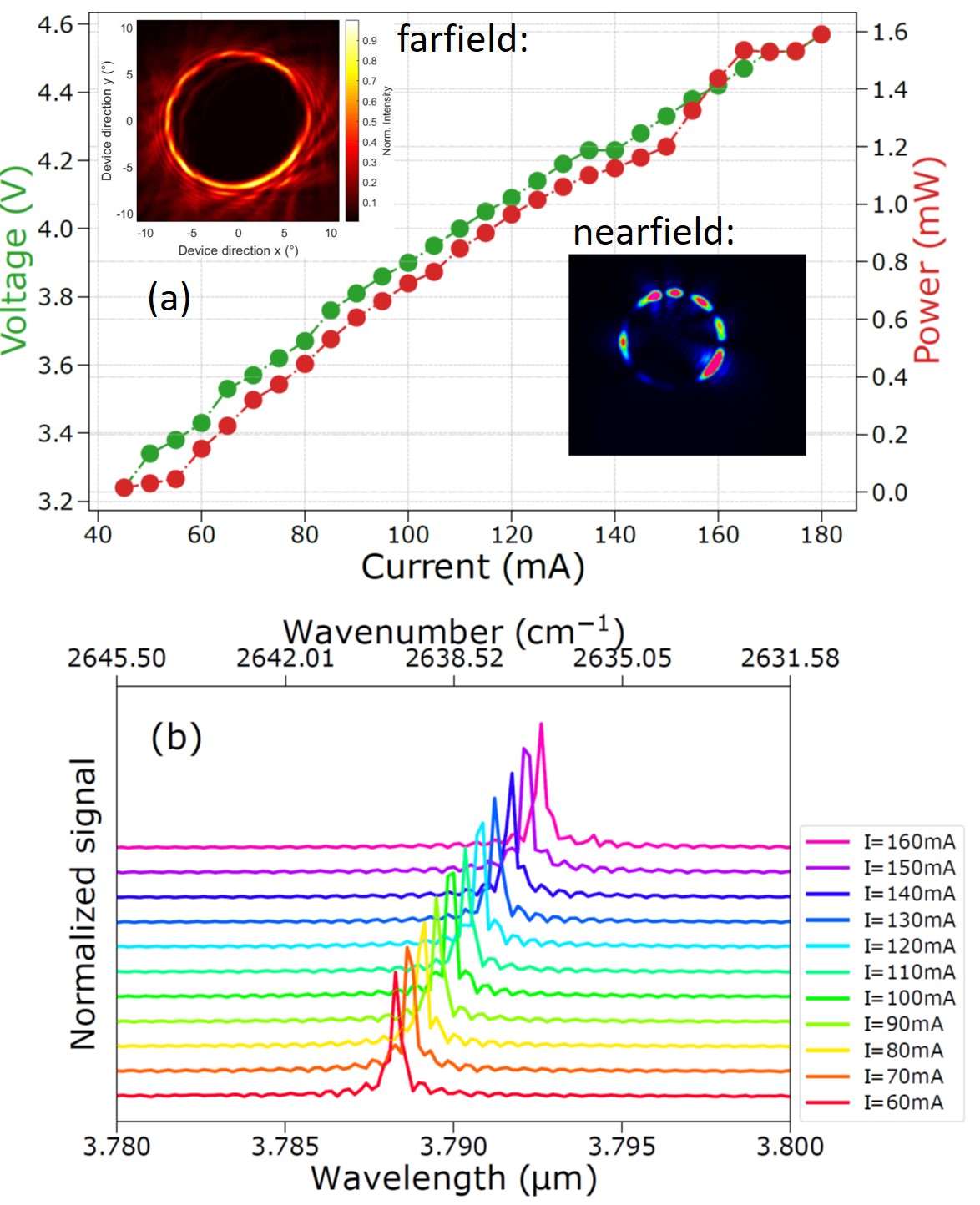}
    \caption{Ring-ICL characterization at a fixed temperature of \SI{16}{\celsius}. (a) LIV curve of the ring-ICL analyzed between $\sim$40~mA and 180~mA. The measured optical power is shown in red and the associated voltage in green. Insets: farfield measured with a HgCdTe-detector on a xy-stage and at a distance of 20~cm to the ring-ICL device and nearfield beam profile obtained with a mid-IR camera (FLIR, ATS SC7000). (b) Corresponding, individually normalized, bias-dependent emission spectra of the ring-ICL measured with an optical spectrum analyzer (FTIR 721, Bristol). 
    The emission spectra show a side-mode suppression ratio of up to 35 dB.}
    \label{fig:LiV-OS}
\end{figure}
First, a typical ring-ICL is characterized in order to determine its optimal working point when operated at a fixed temperature of \SI{16}{\celsius} in CW mode. In our setup, the ring-ICL is driven by an integrated modular controller (ppqSense s.r.l., QubeCL10) including temperature stabilization by using a thermo-electric cooler. Its laser driving unit is characterized by a low bias current noise density, typically around \SI{300}{ pA /\sqrt{\hertz}}, for reducing its effect on the intensity noise of the operated device. As shown by the LIV curve in Fig. \ref{fig:LiV-OS}(a), the tested ring-ICL exhibits a lasing threshold of around \SI{50}{\mA} when operated at \SI{16}{\celsius}, while it reaches its maximum optical output power of approximately \SI{1.6}{\mW} at \SI{160}{\mA}. Regarding the measured optical spectra shown in Fig.~\ref{fig:LiV-OS}(b), the laser maintains a well-defined single-mode emission at around \SI{3.79}{\um} within its whole working range, reaching a side-mode suppression ratio up to 35~dB. As expected, the emission peak moves to longer wavelengths with increasing laser bias current, going from $\mathrm{\lambda}$  = \SI{3.788}{ \um} at \SI{60}{\mA} to $\mathrm{\lambda}$ = \SI{3.793}{ \um} at \SI{160}{\mA}. By analyzing the laser peak emission wavelength as a function of bias current, we obtain a current-tuning coefficient of $\mathcal{T} = (903 \pm 2)$~MHz/mA. More details regarding this analysis are presented in Appendix~\ref{supplsec:tuning-coef}. \\
While the optical emission power of this specific ring-ICL is limited, especially when compared to typical mid-IR QCLs or ICLs, which both can reach emission powers of tens to hundreds of milliwatt~\cite{Kim:15,vurgaftman2013interband,razeghi2009high,bewley2013high}, our ring-device is able to operate at very low consumed electrical power (at maximum bias: $\sim$160~mA at $\sim$4.5~V). This demonstrates its suitability for in-field applications where energy resources are limited to battery operation or even solar energy only.~\cite{wang2017portable,christensen2010thermoelectrically} Moreover, an optical emission power of about \SI{1}{mW} is often sufficient for sensing applications~\cite{vurgaftman:2015}, as long as the target wavelength is precisely hit. Indeed, depending on the detector sensitivity, also hundreds of microwatt of optical power can be sufficient for transmission spectroscopy applications~\cite{Galli:2014a}, as well as novel chip-level applications using compact photonic integrated circuits\cite{Hinkov2022,Pilat2023}. Thus, the high spectral purity and stability of our ring-ICL can be considered suitable for different state-of-the-art applications, including cavity-enhanced spectroscopy experiments \cite{mhanna2020cavity,he2018dual,richard2016optical,manfred2015optical}, free-space optical communication \cite{soibel2009high} and metrological measurements\cite{maddaloni2013laser}. \\
To finalize the basic optical characterization of the ring-ICL, we report, in the inset of Fig. \ref{fig:LiV-OS}, both the acquired nearfield and the farfield profile of the device, measured either with a mid-IR camera (nearfield, FLIR, ATS SC7000) or with a MCT detector mounted on a translational xy-stage which was placed at a distance of 20~cm from the ring-ICL. The ring-shaped geometry with its typical dark central part and a narrow circular beam hosting the device power can clearly be observed in both cases. Here, the nearfield shows some inhomogeneities, likely originating from imperfections in the grating structure, which vanish in the farfield profile due to constructive interference with the photons emitted from other parts of the ring.

\subsection{Intensity noise characterization using balanced detection}

\begin{figure}[!h]
    \centering
    \includegraphics[width=1\columnwidth]{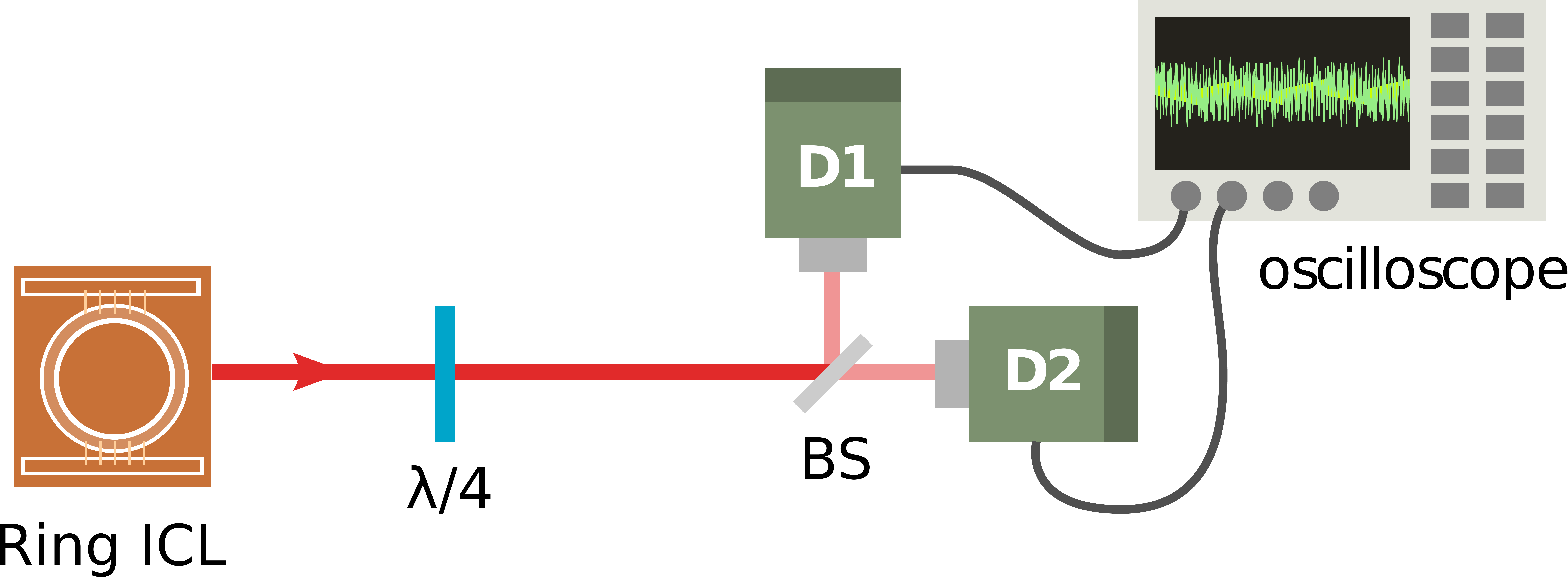}
    \caption{Schematic representation of the experimental setup used for the balanced detection. The figure is re-adapted from \cite{Gabbrielli:2021}.}
    \label{fig:setup}
\end{figure}
In order to understand the intensity noise features of the presented ring-ICL, we analyze its INPSD using a balanced detection experiment. In this setup, sketched in Fig.~\ref{fig:setup}~\cite{Gabbrielli:2021}, the light under investigation is split into two identical beams via a 50/50 beam splitter (BS) and acquired via two commercial \ce{HgCdTe} photovoltaic detectors (D1 and D2) equipped with a 5-MHz-bandwidth pre-amplifier (VIGO Photonics S.A., amplifier: PIP-UC-LS, detector: PV-4TE-4-1x1). The electronic architecture of the detectors consists of two amplifier stages: a 5-k$\mathrm{\Omega}$ pre-amplifier transimpedance stage, where the DC-output is collected, and a second stage for which coupling (i.e. AC or DC) and gain can be chosen via a PC software (VIGO Photonics S.A., Smart Manager). In our case, the AC-coupled second stage is used to amplify the AC-voltage noise amplitude by a factor of 85~V/V~\cite{note1}. The detectors are maintained at a fixed temperature of T~=~200~K by a four-stage-Peltier cooling system using a thermometric cooler controller (VIGO Photonics S.A., PTCC-01-BAS).  The signals are analyzed in the time domain. In particular, a 12-bit oscilloscope (Tektronix, MSO64) is used to acquire the two detectors' output signals in a 20-ms time window and at a fixed sampling rate of \SI{31.25}{Ms/\second}. In our measurements, the oscilloscope bandwidth is limited to \SI{20}{MHz}. Finally, a Python script is used to compute the sum and the difference of the acquired signal and to convert them from the time to the frequency domain, computing the INPSD of the difference and of the sum~\cite{Gabbrielli:2021}. Since at each point the polarization is tangential to the beam (i.e. circular), a $\mathrm{\lambda}$/4 wave-plate is placed just after the device to retrieve a linear polarization~\cite{note2}. The emitted laser beam from the chip is uncollimated, therefore, we placed an additional 50-mm lens in front of each detector to collect all the light within its 1$\times$1 mm$^2$ collection area. \\
For evaluating the performance of the assembled balanced detection system, we performed a preliminary characterization of the used photodetectors. One key parameter in our measurements is the detector responsivity, defined as the detector output signal (voltage or current) as a function of the incident optical power. In particular, in order to perform a balanced detection where the common noise of both arms is suppressed at the shot-noise level, it is necessary to use two photovoltaic detectors with a responsivity that is as similar as possible. Otherwise, the detection is unbalanced in favor of one of the two arms, even when investigating two initially identical incident optical signals on D1 and D2. Thus, when performing the balanced detection experiments in the linear responsivity regime of two detectors, the INPSD computed from the difference of the photocurrent output signals is expected to be at the shot-noise level and, therefore, proportional to the incident power impinging on the BS~\cite{Gabbrielli:2021}. This is true in the limit given by the maximum common mode rejection ratio (CMRR) achievable with our setup, i.e. the maximum excess of noise with respect to the shot-noise level that can be canceled with our differential measurement~\cite{Gabbrielli:2021}. Instead, the sum of the two photocurrent AC output signals corresponds to the measurement of the whole intensity noise associated with the radiation impinging on the balanced detector. It is linked to the intensity noise of the laser minus a possible attenuation factor (due to the losses experienced by the propagating beam and the detector efficiency), plus an extra contribution due to the coupling of the tested radiation with the vacuum field caused by the losses~\cite{Gabbrielli:2021}. Under the condition of balanced detection performed in the linear responsivity regime, and assuming the noise level does not exceed the maximum CMRR, it is sufficient to directly compare the retrieved INPSD of sum and difference for judging whether the light collected from the source under investigation is shot-noise-limited. This means that its photons are Poissonian-like distributed, as expected for a coherent light source~\cite{loudon:2000book}. Based on these considerations, we carefully selected two photodetectors with a very similar responsivity at $\mathrm{\lambda}$~=~$\SI{3.79}{\micro m}$ of $\mathrm{R_1}$~=~(0.422$\pm$0.004)~\SI{}{A/W} and $\mathrm{R_2}$~=~(0.396$\pm$0.006)~\SI{}{A/W}, respectively. Furthermore, when the differential measurement is performed, a CMRR up to \SI{25}{dB} is achievable in the tested bandwidth. More details of this analysis are available in Appendix \ref{supplsec:CMRR}.\\
Fig. \ref{fig:INPSD_spectra} shows the INPSD of the ring-ICL analyzed at 140 mA, which corresponds to a condition in which the laser is neither affected by noise contributions from spontaneous emission events close to the laser threshold nor by any saturation effects close to the device rollover, as shown in Fig. \ref{fig:LiV-OS}. The output power under these driving conditions is around \SI{1.2}{mW}. Therefore the detectors receiving each around \SI{0.6}{mW}, are not saturated (the optical losses due to the optical tools, e.g mirrors, waveplate, lenses are around 2\%). As evidenced in Fig. \ref{fig:INPSD_spectra}, the INPSD of the difference signal (blue trace) corresponds to a direct measurement of the shot-noise level: indeed, the INPSD of the difference signal overlaps with the red trace, which shows the sum of the background noise (gray trace) and the theoretically computed shot-noise power spectral density (PSD) (dashed black line). To retrieve this latter quantity, we measure the DC output of the two photovoltaic detectors and calculate the shot-noise PSD as PSD$_{\mathrm{SN}}=2 e(V_1+V_2)/R$, where $e$ is the electron charge, $V_{1,2}$ are the voltages measured at the two first-stage transimpedance DC-outputs of both detectors and $R$ is the transimpedance resistance value. The red trace is then displayed as the sum of the gray and dashed black trace, to take into account the effect of the background with respect to the calculated shot-noise level. It is important to note that, despite a non-negligible contribution of the background in the measured shot-noise, the INPSD of the difference signal lies well above the sole detector background level, reaching a so-called clearance, defined as the ratio between the INPSD of the difference signal and the detector background, of up to 6~dB at a Fourier frequency of about 1~MHz\cite{Gabbrielli:2021}. This result confirms the possibility of performing shot-noise limited detection with the assembled setup, e.g. the setup can be successfully applied in a homodyne detection scheme using the tested ring-ICL as a local oscillator~\cite{Gabbrielli:2021}. With this purpose, the optimal working conditions are those which guarantee to exploit a clearance as high as possible to minimize the effect of the background on the measurement and thus potentially increase the possibility of exploring sub-shot-noise signal levels in balanced detection~\cite{Gabbrielli:2021,loudon:2000book}. In our case, the best working conditions are therefore the use of the ring-ICL at a driving current of \SI{140}{mA} where it emits a power of \SI{>1}{mW} which allows reaching the best clearance (i.e. \SI{6}{dB}) with the assembled setup. In view of possible non-classical application, one major limitation arises from the limited quantum efficiency of the detectors (i.e. the number of generated electrons in a detector as a function of the number of impinging photons). As shown in the supplementary material~\ref{supplsec:detresp} this quantity lies around 13-14~\% at the investigated wavelength. Still, the here presented results give a good starting point for the development of future quantum technology systems based on the light source tested in this work. Next, we will seek to implement commercial detectors, optimized for working in the \SI{4}{\micro m} window, with higher quantum efficiency, to potentially address quantum optics applications, where losses directly correspond to a degradation of the non-classicality of a tested non-classical signal (e.g. a squeezed state of light characterized by sub-shot-noise level amplitude noise). This is done by mixing it with the vacuum state of the electromagnetic field for a percentage corresponding to the amount of the losses~\cite{loudon:2000book, Gabbrielli:2021}.

\begin{figure}[!h]
    \centering
    \includegraphics[width = 1 \columnwidth]{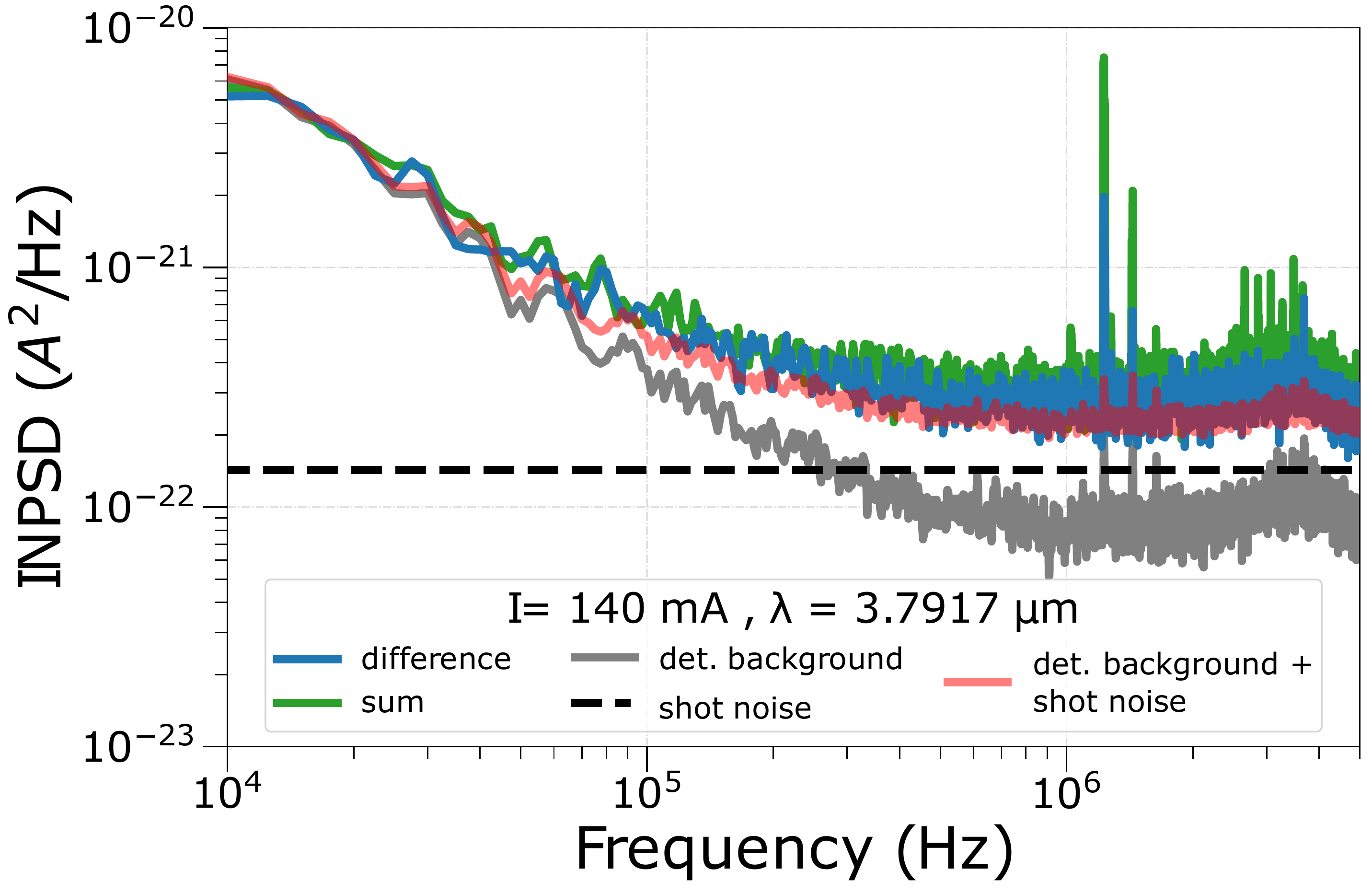}
    \caption{Ring-ICL Intensity Noise Power Spectral Density (INPSD) analysis performed at a fixed temperature of \SI{16}{\celsius} and at a laser bias current of 140 mA. The INPSD sum and difference signal  traces are shown in green and blue, respectively. The dashed black line represents the theoretical shot noise level, obtained from the DC outputs of the detectors, while the detector background is shown in gray. The red line shows the sum of the shot-noise level and detector background noise. In the INPSD of the detector background and of the laser, some spurious noise peaks at slightly above \SI{1}{MHz} are present. They are due to technical noise originating from different sources including intrinsic electronic noise of the current driver, mass-loop noise due to its power supply, and the supply used for the detectors. This technical noise can be reduced e.g. by using battery operation. It is important to note, that even though these peaks are present, still a shot-noise limited intensity noise for the tested ICL is demonstrated, with the exception of those few particular frequencies.}
    \label{fig:INPSD_spectra}
\end{figure}
\begin{figure}[!h]
    \centering
    \includegraphics[width = 1 \columnwidth]{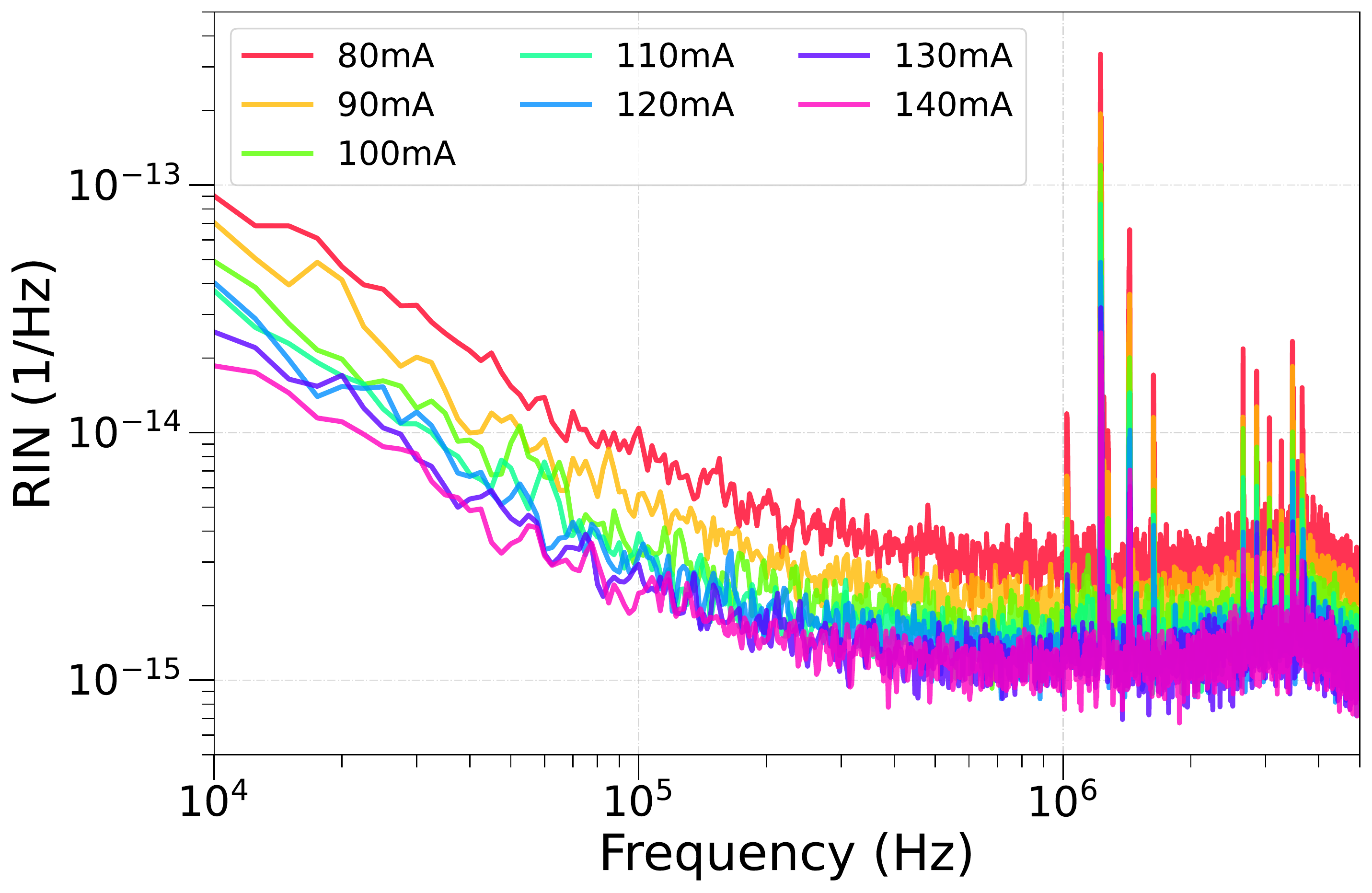}
    \caption{Relative Intensity Noise (RIN) of the ring-ICL measured for different bias currents at a fixed temperature of \SI{16}{\celsius}. As already discussed in Fig. \ref{fig:INPSD_spectra}, also the here-reported RIN shows some spurious noise peaks due to the presence of excess technical noise, originating from the laser driving unit.}
    \label{fig:RIN_spectra}
\end{figure}

Coming back to the characterization of the tested ICL, it clearly benefits from a shot-noise-limited intensity noise within the tested detector efficiency. Indeed, the obtained data shows an INPSD of the sum signal (green trace, Fig.~\ref{fig:INPSDs}) that is superimposed with the INPSD of the difference trace in blue for the entire investigated Fourier frequency domain. In Appendix~\ref{supplsec:Diff_curr_INPSD} we also demonstrate that this interesting behavior is similar for different laser drive currents, at fixed laser temperature. The shot-noise limited operation represents an important feature in ring-ICLs for applications requiring a well-suppressed-intensity-noise light source, such as in quantum homodyne detection~\cite{Gabbrielli:2021,loudon:2000book}, high-sensitivity interferometry~\cite{plick2010coherent,abbott2009ligo} and spectroscopy~\cite{truong2012absolute}. With this purpose, it is worth noting that at lower frequencies (up to \SI{100}{KHz}) all traces are background noise limited. Therefore, in view of future applications, the optimum working range for our balanced detection setup is in the frequency range between \SI{100}{kHz} and \SI{5}{MHz}, where there is the roll-over due to the limited bandwidth of the detectors. \\
Finally, Fig. \ref{fig:RIN_spectra} depicts the RIN of the ring-ICL at different bias currents for a fixed temperature of \SI{16}{\celsius}. The RIN is defined as the INPSD of the sum signal normalized to the square of the sum of the photo-currents measured by the two photo-detectors. As expected, the RIN decreases with increasing laser bias current for measurements between I = \SI{80}{\mA} and I = \SI{140}{\mA}. 

\section{Conclusion}

In conclusion, we investigated the noise characteristics of a 2\textsuperscript{nd}-order DFB ring-ICL emitting at $\lambda = \SI{3.79}{\um}$ for a fixed temperature of \SI{16}{\celsius}. The INPSD level found at a driving current I~=~140~mA, i.e. far from the laser threshold and from the laser rollover, with a balanced-detection setup demonstrates shot-noise limited operation between \SI{10}{kHz} and \SI{5}{MHz}. In the setup, we employed two HgCdTe photovoltaic detectors with similar responsivity, which are moreover linear in the whole range of investigated laser bias currents. Sub-shot-noise detection is shown to be potentially possible with such a configuration. For this purpose, the detector quantum efficiency should be improved, in order to enhance the chance of unveiling sub-classical emission, by limiting losses. \\
We further investigated the RIN of our experimental configuration, obtaining decreasing RIN values with increasing laser bias currents. Moreover, in contrast to previous RIN studies in ICLs~\cite{Deng:19}, we show that our ring-DFB laser exhibits orders of magnitude lower values for all frequencies up to \SI{5}{MHz}. \\
In the future, better detection technology with significantly higher quantum efficiencies (currently $\sim$13--14~\%) are needed, to explore the sub-classical regime and quantum optics applications.

\begin{acknowledgments}
The authors gratefully thank M. Schinnerl for expert technical assistance in focused ion beam (FIB) cutting and SEM characterization of the devices. Fruitful discussions with W. Schrenk and E. Gornik are greatly acknowledged. \\
The authors acknowledge financial support by the European Union’s Next Generation EU Programme with the I-PHOQS Infrastructure [IR0000016, ID D2B8D520, CUP B53C22001750006] "Integrated infrastructure initiative in Photonic and Quantum Sciences'', by the European Union’s Research and Innovation Programmes Horizon 2020 and Horizon Europe with the cFlow Project [G.A.~n.~828893] "Coherent ultrafast long wave infrared communications", the MUQUABIS Project [G.A.~n.~101070546] "Multiscale quantum bio-imaging and spectroscopy", by the European Union’s QuantERA II [G.A.~n.~101017733] -- QATACOMB Project "Quantum correlations in terahertz QCL combs'', by Fondazione CR Firenze through the SALUS project, and by the Italian ESFRI Roadmap (Extreme Light Infrastructure -- ELI Project). The authors are grateful for financial support by the Austrian Research Promotion Agency (FFG) through the ATMO-SENSE project [G.A.~n.~1516332] "Novel portable, ultra-sensitive, fast and rugged trace gas sensor for atmospheric research based on photothermal interferometry" and the NanoWaterSense project [G.A.~n.~873057] "Mid-IR Sensor for Trace Water Detection in Organic Solvents, Oils and Petrochemical Products". Financial support by the State of Bavaria is grately acknowledged.

\end{acknowledgments}

\section*{Data Availability Statement}

The data that support the findings of this study are available from the corresponding author upon reasonable request.

\appendix

\section{Tuning coefficient characterization}
\label{supplsec:tuning-coef}

\begin{figure}[!h]
    \centering
     \includegraphics[width = 1\columnwidth]{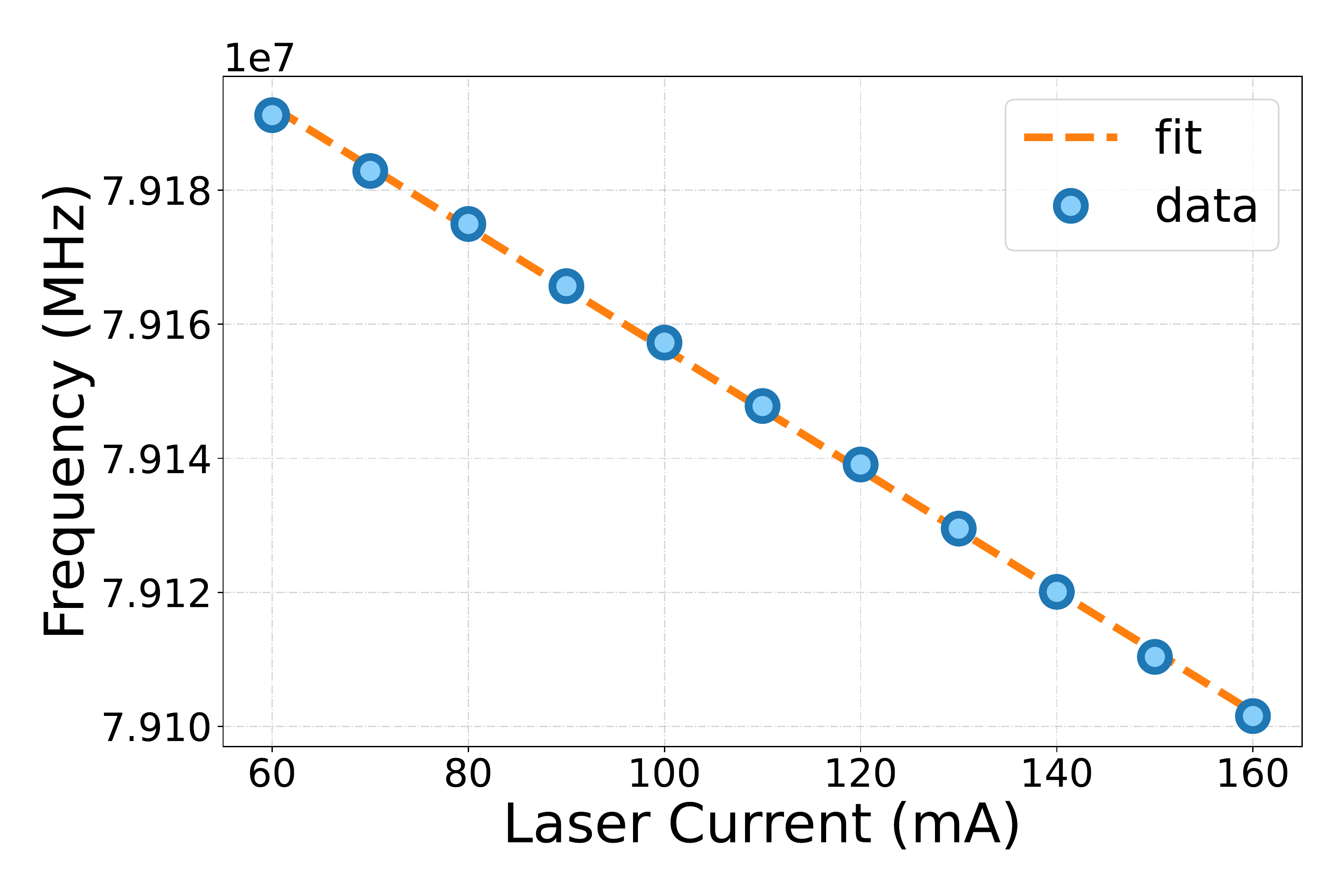}
    \caption{Current-Tuning coefficient characterization. The emission wavelength peaks are first converted into the frequency domain and then plotted as a function of the laser current (blue circles). The linear fit (dashed orange line) procedure allows us to estimate the tuning coefficient $\mathcal{T}$ = (903 $\pm$ 2) MHz/mA.}
    \label{fig:tuningcoeff}
\end{figure}

The estimation of the current-tuning coefficient gives insight into the relation between the emission wavelength of a laser and the applied bias current and was experimentally demonstrated to be pretty much linear \cite{du2016dynamic,hou2012method}. For each optical spectrum shown in Fig.~\ref{fig:LiV-OS}(b) the corresponding emission peak is fit to the following Gaussian function: 
\begin{equation}\label{eq:1}
f(\lambda) = A \mathrm{e}^{-4 \mathrm{ln}(2)(\lambda-\lambda_0)^2/\Delta\lambda^2}+f_0
\end{equation}
where $A$ is the amplitude, $\lambda_\mathrm{0}$ the wavelength of the emission peak, $\Delta\lambda$ the full width at half maximum (FWHM) of the peak, and $\mathrm{f_0}$ an offset. All of them are free parameters in the fit procedure. After converting the wavelength to the frequency domain by using the relation $\mathrm{c}$ = $\mathrm{ f \cdot \lambda}$, their linear regression is computed (see Fig.~\ref{fig:tuningcoeff}) to extract the current-tuning coefficient of our ring-ICL $\mathcal{T} = (903 \pm 2)$~MHz/mA. 

\section{Detector Responsivities}
\label{supplsec:detresp}
\begin{figure}[!h]
    \centering
     \includegraphics[width=1\columnwidth]{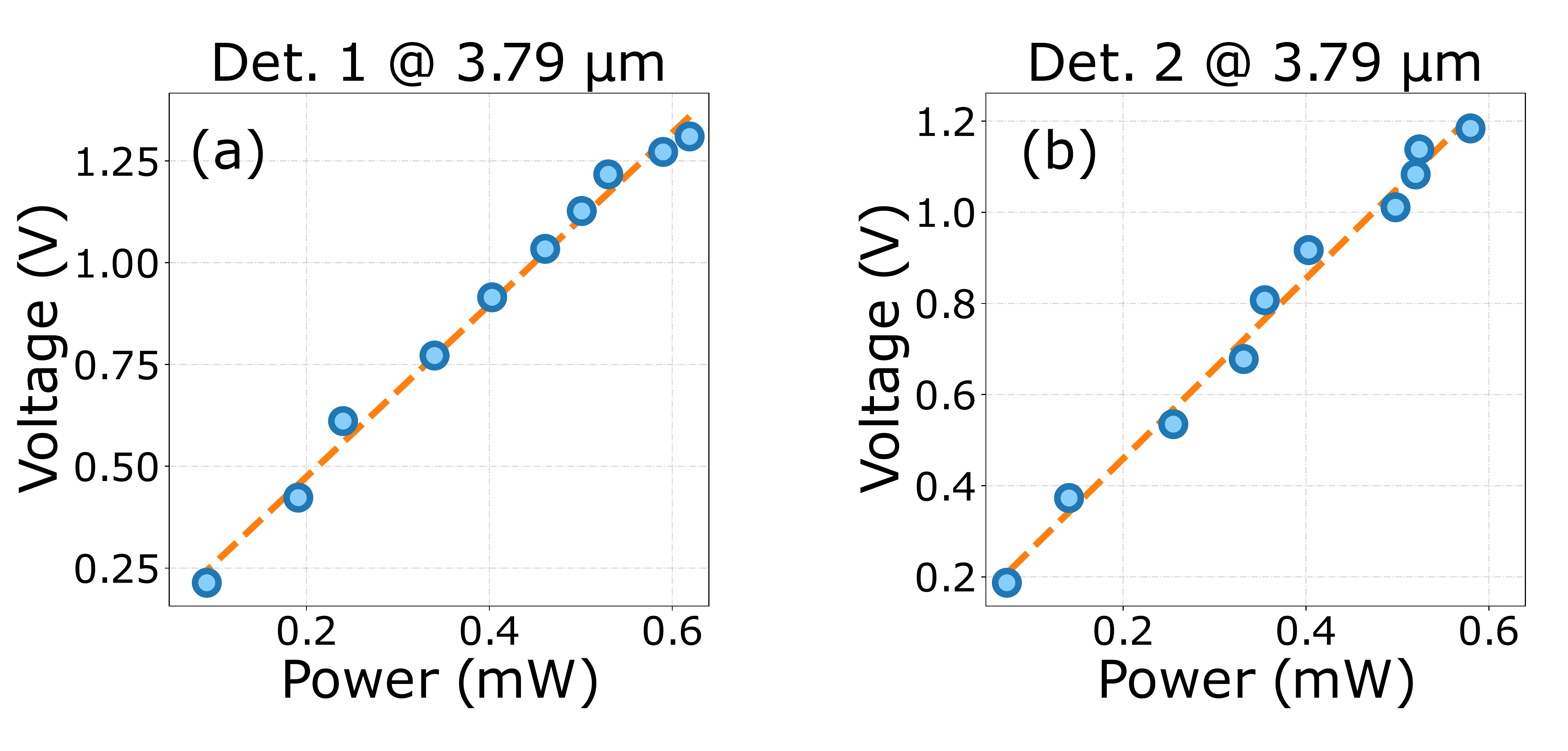}
    \caption{Responsivity of both detectors measured at a wavelength of $\mathrm{\lambda}$~=~\SI{3.79}{\micro m} using the ring-ICL. A linear fit (dashed orange line) is computed to extract the following responsivities: (a) $\mathrm{R1}$~=~(0.422$\pm$0.004)~A/W and (b) $\mathrm{R2}$~=~(0.396$\pm$0.006)~A/W.
    }
    \label{fig:responsivity}
\end{figure}

The responsivity quantifies the efficiency of the detector for converting an input optical power signal into an output photo-current (or photo-voltage). In the limit of linear response, it is given by:
\begin{equation}\label{eq:2}
\mathrm{R = \eta_{qe}{\frac{\lambda e}{hc}}} 
\end{equation}
where $\mathrm{\eta_{qe}}$ is the quantum efficiency, $e$ the electron charge, $\lambda$ the wavelength, $h$ the Planck constant, and $c$ the speed of light. 
In our work, we measure the responsivity of both detectors. At a fixed laser temperature of \SI{16}{\celsius} we span the bias current from 60~mA to 150~mA with a step size of 10 mA. For each current value,  we measure the emission power in front of the detector via a  power meter (Thorlabs, PM400) and the pre-amplifier first-stage DC-output voltage. The acquired data is then fitted with a linear regression, and the estimated curve slope (in [V/W]) is divided by the transimpedance resistance (5 k$\mathrm{\Omega}$ in our case) to obtain the detector responsivity in terms of the photocurrent. \\
As shown in Fig.~\ref{fig:responsivity}, for a wavelength of $\lambda$~=~\SI{3.79}{\micro m}, the responsivity of Detector 1 and Detector 2 is $\mathrm{R1}$ = (0.422$\pm$0.004)~A/W and $\mathrm{R1}$ = (0.396$\pm$0.006)~A/W, respectively. From equation \ref{eq:2}, it is possible to extract a quantum efficiency $\mathrm{\eta_{qe}}$ of (0.138$\pm$0.001) for the first detector and of (0.129$\pm$0.002) for the second one. 

\section{Detector CMRR}
\label{supplsec:CMRR}
\begin{figure}[!h]
    \centering
    \includegraphics[width=1\columnwidth]{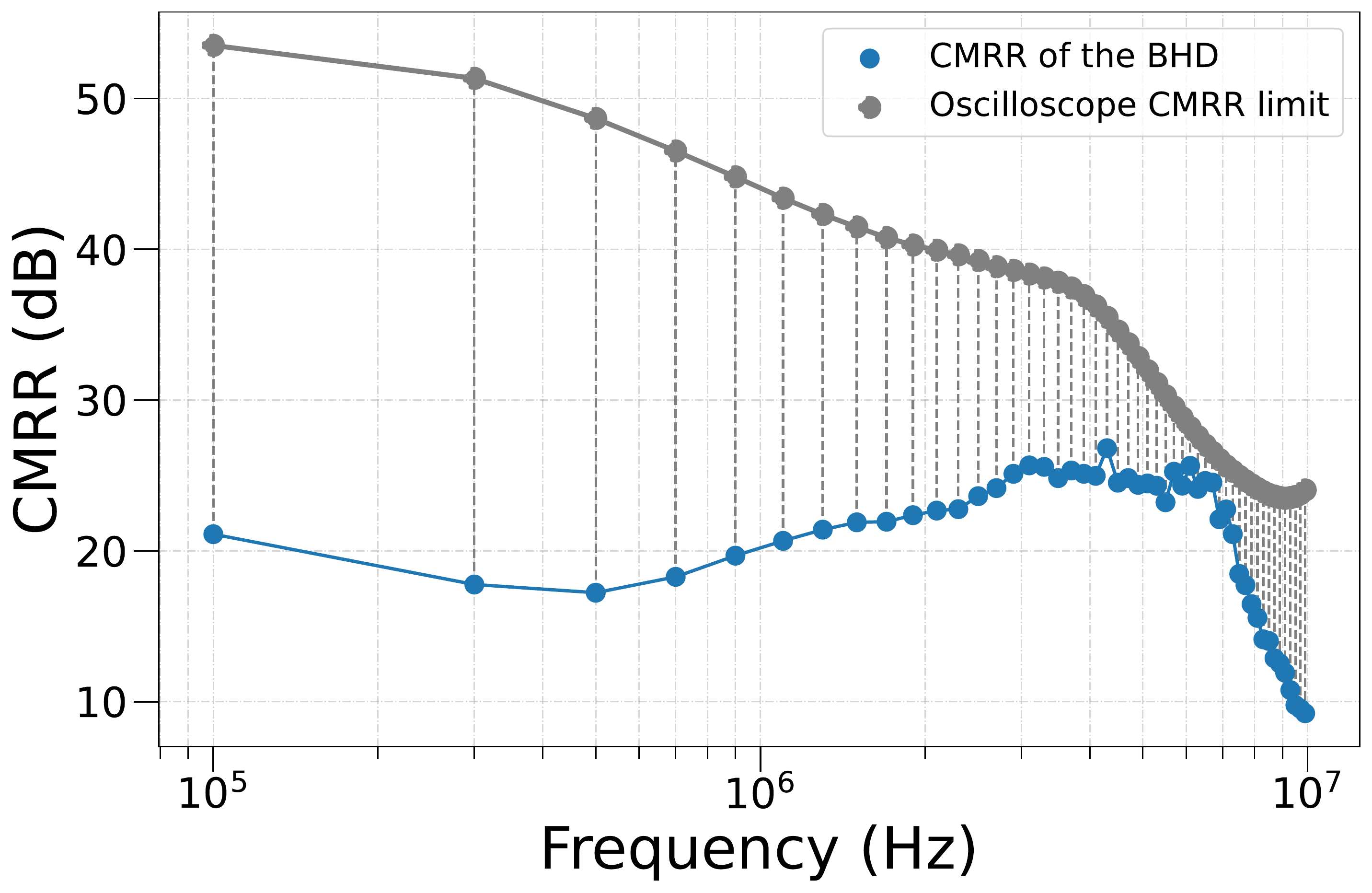}
    \caption{Common-Mode Rejection Ratio (CMRR) of the balanced detector (blue circles) and of the oscilloscope (grey circles).}
    \label{fig:CMRR}
\end{figure}

 The CMRR is calculated as the ratio between the INPSD of the sum and of the difference and quantitatively assets the noise rejection capability of our balanced detector \cite{Gabbrielli:2021}. To measure the maximum CMRR achievable with our setup, we modulate the laser current via the current driver module. The used modulation signal is a square wave of amplitude \SI{0.8}{V} and with a carrier frequency of \SI{100}{kHz}. The FFT of a square signal is characterized by all the odd harmonics generated starting from the main frequency.  As shown in Fig. \ref{fig:CMRR}, the CMRR is calculated in this measurement at the harmonic frequencies of the used \SI{100}{kHz} square wave. 
The experimental data obtained with this procedure (blue dots) is compared to the maximum CMRR achievable with the oscilloscope (grey dots) when the same square signal is  split via a t-connector and sent directly onto the two oscilloscope channels used for the measurements, as described in \cite{Gabbrielli:2021}. With our setup, we are able to achieve up to \SI{25}{dB} (e.g. around \SI{3}{MHz}).

\section{INPSD analysis at different laser currents}
\label{supplsec:Diff_curr_INPSD}
 \begin{figure}[!h]
    \centering
     \includegraphics[width=\columnwidth]{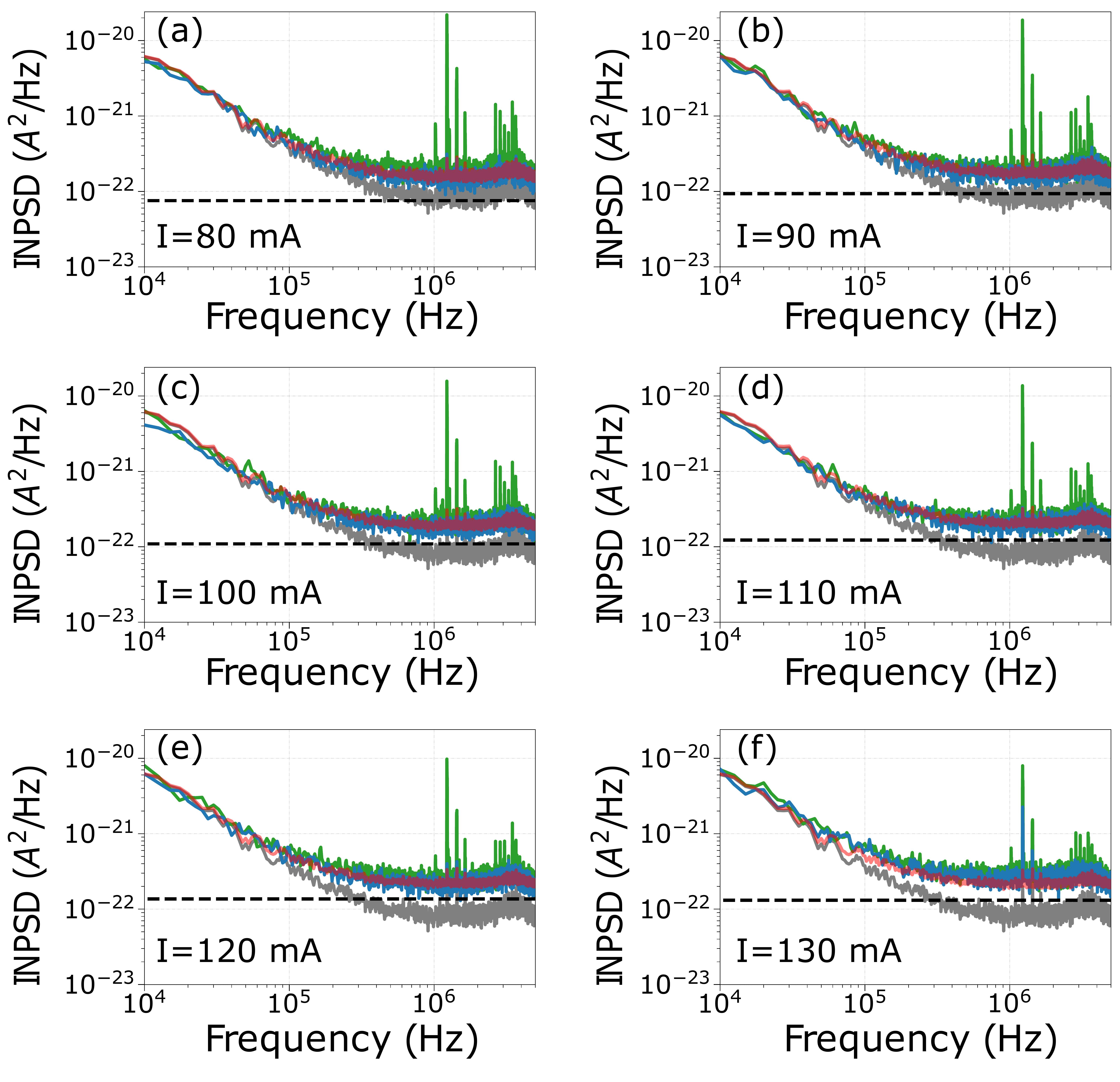}
    \caption{Ring-ICL INPSD analysis at a fixed temperature of 16~$^{\circ}$C and for different laser bias currents: (a) I~=~80~mA, (b) I~=~90~mA, (c) I~=~100~mA and (d) I~=~110~mA, (e) I~=~120~mA, (f) I~=~130~mA. For each plot the theoretical shot-noise level is shown with a dashed black line, the detector background is depicted in gray, the sum of these two quantities is in red, the INPSD of the difference in blue, and the INPSD of the sum in green. As described in Fig.\ref{fig:INPSD_spectra}, also in these INPSD measurements the spurious noise around 1~MHz is present due to technical noise of a mass loop between the power supply of the detector and of the laser.}
    \label{fig:INPSDs}
\end{figure}
Fig.~\ref{fig:INPSDs} shows the INPSD analysis performed at \SI{16}{\celsius} for different values of the current, i.e. (a) I~=~80~mA, (b) I~=~90~mA, (c) I~=~100~mA, (d) I~=~110~mA, (e) I~=~120~mA, and (f) I~=~130~mA. As expected, the INPSD of the sum and of the difference continuously move away from the detector background, since the shot noise level linearly increases with the photocurrent. In fact, at I = \SI{80}{mA} (Fig.~\ref{fig:INPSDs}(a)), where the laser power is only 0.45 mW, the noise spectra slightly exceed the detector background, while at I = 130 mA (Fig.~\ref{fig:INPSDs}(f)), where the laser reaches an emission's power equal to 1.1 mW, the difference between the two levels is more appreciable, reaching a clearance of up to \SI{6}{dB} at a Fourier frequency of around \SI{1}{MHz}~\cite{Gabbrielli:2021}.
The ring-ICL intensity noise remains at the shot noise level within the whole range of investigation.

\bibliography{references,Ring-ICL-Noise}

\begin{thebibliography}{64}
\bibitem [{not({\natexlab{a}})}]{note1}%
  \BibitemOpen
  \href@noop {} {} ({\natexlab{a}}),\ \bibinfo {note} {via the Smart Manager
  control platform we fixed the gain at 20, but we also measure a minimum
  amplification amount between the 1 stage output voltage and the 2st stage
  output voltage equal to 4.25 V/V when the AC gain is set at 1. The two factor
  ( e.g 4.25 x 20 in our case) need to be multiplyed to retrieve the actual
  voltage amplifaction factor.}\BibitemShut {Stop}%
\bibitem [{not({\natexlab{b}})}]{note2}%
  \BibitemOpen
  \href@noop {} {} ({\natexlab{b}}),\ \bibinfo {note} {a $\mathrm{\lambda}$/2
  wave-plate is also needed to equally balance the two split arms}\BibitemShut
  {NoStop}%
\end{thebibliography}

\end{document}